\documentclass[prl,twocolumn,showpacs,amsmath,amssymb,superscriptaddress]{revtex4}
\usepackage{graphicx}
\usepackage{dcolumn}
\usepackage{bm}
\usepackage{epsfig}
\begin{document}


\title{Magnetic order in the quasi-one-dimensional spin 1/2 chain, copper pyrazine dinitrate}

\author{T. Lancaster}
\email{t.lancaster1@physics.ox.ac.uk}
\author{S.J. Blundell}
\author{M.L. Brooks}
\author{P.J. Baker}
\affiliation{
Clarendon Laboratory, Oxford University Department of Physics, Parks
Road, Oxford, OX1 3PU, UK
}
\author{F.L. Pratt}
\affiliation{
ISIS Facility, Rutherford Appleton Laboratory, Chilton, Oxfordshire OX11 0QX, UK}
\author{J.L. Manson}
\affiliation{Department of Chemistry and Biochemistry,
Eastern Washington University,
Cheney, WA 99004, USA
}
\author{C.P. Landee}
\affiliation{Department of Physics, Clark
University, Worcester, Massachusetts 01610, USA}

\author{C. Baines}
\affiliation{Swiss Muon Source, Paul Scherrer Intitut, CH-5253, PSI Villigen, Switzerland}
\date{\today}

\begin{abstract}
We present the first evidence of magnetic order in the
quasi-one-dimensional spin 1/2 molecular chain compound, 
copper pyrazine dinitrate Cu(C$_{4}$H$_{4}$N$_{2}$)(NO$_{3}$)$_{2}$. 
Zero field muon-spin relaxation
measurements made
at dilution refrigerator temperatures show oscillations in the measured asymmetry,
characteristic of a quasistatic magnetic field at the muon
sites. Our measurements provide convincing evidence for long
range magnetic
order below a temperature $T_{\mathrm{N}}=107(1)$~mK. This leads to an 
estimate of the interchain
coupling constant of $|J'|/k_{\mathrm{B}}=0.046$~K and 
to a ratio $|J'/J| = 4.4 \times 10^{-3}$. 
\end{abstract}

\maketitle

\begin{figure*}
\begin{center}
\epsfig{file=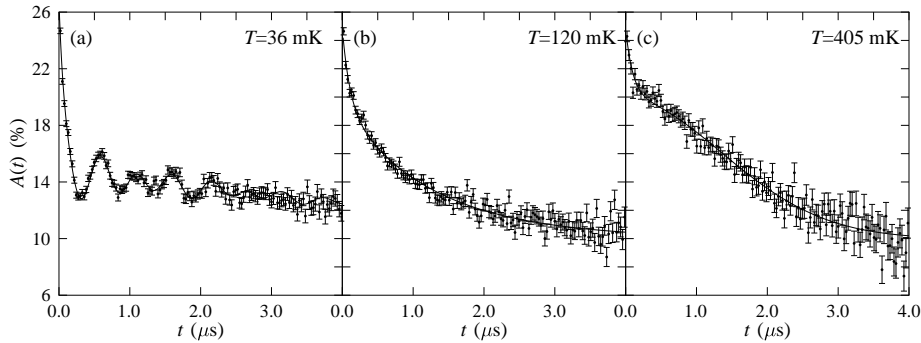,width=13.5cm}
\caption{ZF $\mu^{+}$SR spectra for CuPzN at three 
temperatures. (a)  
$T=36$~mK: Oscillations are observed in the asymmetry, characteristic of
long-range magnetic order. A fit is shown to Eq.(\ref{fitfunc}).
(b) $T=120$~mK: the spectra are described by two components. A fast
relaxing signal, probably due to the existence of a radical state, and 
a larger component, seen to be exponential in form. 
A fit is shown to Eq.(\ref{hightemp}). (c) $T=405$~mK: The
fast relaxing component is still evident, although the larger component
now shows a Gaussian form. A fit is shown to Eq.(\ref{hightemp}).
\label{data}}
\end{center}
\end{figure*}

Experimental studies of quantum many-body physics rely on the existence
of simple and well characterised model systems. One such system 
is the linear chain
Heisenberg antiferromagnet (LCHAFM) which is described by the Hamiltonian 
\begin{equation}
H=\sum_{i}[-J \mathbf{S}_{i} \mathbf{S}_{i+1} +g \mu_{\mathrm{B}} \mathbf{B}
\cdot \mathbf{S}_{i}],
\end{equation}
 where
$J$ is the nearest neighbour exchange constant along the
chain and $\mathbf{B}$ is an applied magnetic field. 
For $\mathbf{B}=0$, the system will not show long-range magnetic order
(LRO) for $T>0$, but will display excitations in the form of
$S=1/2$ spinons which form a gapless, two-particle continuum \cite{muller,karbach}.
The magnetic field $\mathbf{B}$ can tune the system through a 
quantum critical point (QCP) \cite{sachdev}, 
(i.e.\ a zero temperature phase transition) at 
$|\mathbf{B}|=|2J/g \mu_{\mathrm{B}}|$ where the spin state becomes fully
polarized \cite{bogoliubov,fledderjohann}. 

Experimental realisations of the LCHAFM have been made in several
copper
chain compounds, the most successful of which is copper pyrazine
dinitrate Cu(C$_{4}$H$_{4}$N$_{2}$)(NO$_{3}$)$_{2}$ or CuPzN. This material
has a relatively low value of $J$($=10.3(1)$~K, determined from high field
magnetization and specific heat measurements \cite{hammar}),
ensuring the QCP is at an experimentally accessible magnetic field,
but also a small ratio of intrachain to interchain exchange constants
(estimated as $|J'/J|<10^{-4}$ \cite{villain}), showing that the chains
are well isolated. This has allowed the magnetic field dependence of
the excitations in the two-spinon continuum to be measured and led to
the identification of an extended critical state \cite{stone},
confirming long-standing predictions of theoretical and numerical
studies \cite{lefmann,karbach2,karbach3}. 
Three-dimensional (3D) LRO has not been detected in CuPzN
down to 70~mK on the basis of specific heat measurements
\cite{mennenga}. However, as we shall demonstrate, 
implanted positive muons are a much more sensitive probe of 3D LRO
than specific heat for the case of very anisotropic magnetic chains. 
In this letter we show for the first time that CuPzN does indeed 
undergo a transition into a state of 3D LRO and provide an estimate 
of $|J'/J|$ which is larger than previously thought.

Zero-field muon-spin relaxation (ZF $\mu^{+}$SR) measurements were made
on 
a powder
sample of CuPzN using the LTF instrument
at the Swiss Muon Source (S$\mu$S), Paul Scherrer Institut, Villigen, 
Switzerland and the MuSR instrument at the
ISIS facility, Rutherford Appleton Laboratory, UK. 
In a $\mu^{+}$SR experiment \cite{steve} spin polarised muons
are implanted into the sample. The quantity of interest
is the asymmetry $A(t)$, which is proportional to the
spin polarization of the muon ensemble.

Example ZF $\mu^{+}$SR spectra measured using a dilution
refrigerator at S$\mu$S
are shown in Fig.\ref{data}. In spectra measured below a temperature 
$T=110$~mK oscillations in the asymmetry are observed
(Fig.\ref{data}(a)). These oscillations are
characteristic of a quasi-static local magnetic field at the 
muon stopping site, which causes a coherent precession of the
spins of those muons with a component of their spin polarization
perpendicular to this local field (expected to be 2/3 of the total
polarization). The frequency of the oscillations is given by
$\nu_{i} = \gamma_{\mu} B_{i}/2 \pi$, where $\gamma_{\mu}$ is the muon
gyromagnetic ratio ($=2 \pi \times 135.5$~MHz T$^{-1}$) and $B_{i}$
is the average magnitude of the local magnetic field at the $i$th muon
site. Any fluctuation in magnitude of these local fields will
result in a relaxation of the oscillating signal, described by
relaxation rates $\lambda_{i}$. The presence of oscillations at
low temperatures in CuPzN provides convincing evidence
that this material is magnetically ordered below 110~mK.

 Two separate
frequencies were identified in the low temperature spectra,
corresponding to two magnetically inequivalent muon stopping sites in the
material. We note here that two muon sites have been observed in other
copper
pyrazine-based  compounds \cite{tom}.
The precession frequencies, which are proportional to the internal
magnetic field as experienced by the muon, 
 may be viewed as an effective order parameter
for these systems. 
In order to extract the temperature dependence of the frequencies, 
the low temperature data were fitted to the functional form
\begin{eqnarray}
A(t) &=& A_{1} \exp (-\lambda_{1} t )\cos (2 \pi \nu_{1} t + \phi_{1})
\nonumber
\\
& &      +A_{2} \exp (-\lambda_{2} t )\cos (2 \pi \nu_{2} t + \phi_{2})  \nonumber \\
& &      +A_{3} \exp(-\Lambda t)
         +A_{\mathrm{bg}}\exp(-\lambda_{\mathrm{bg}} t)\label{fitfunc}, 
\end{eqnarray}
where $A_{3} \exp(-\Lambda t)$ accounts for the contribution
from those muons with a spin component parallel to the local magnetic
field. The term $A_{\mathrm{bg}}\exp(-\lambda_{\mathrm{bg}} t)$ reflects the
small signal from those muons which stop in the silver sample holder
or cryostat tail (with $\lambda_{\mathrm{bg}} \ll \Lambda$). 

\begin{figure}
\begin{center}
\epsfig{file=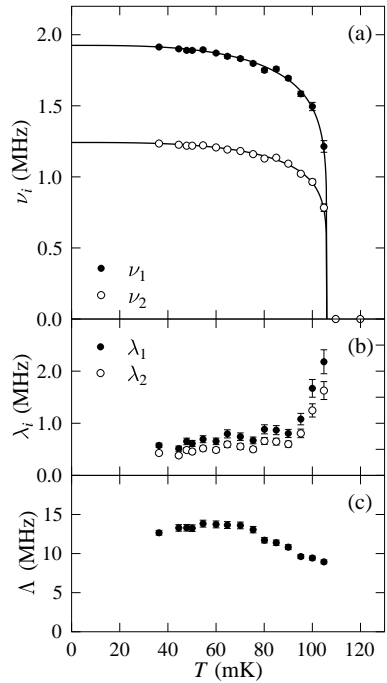,width=6.0cm}
\caption{
Temperature evolution of the variable parameters in Eq.(\ref{fitfunc}) 
for $T \leq 140$~mK.
(a) Muon precession frequencies $\nu_{i}$ with fits shown to the expression
$\nu_{i}(T) =\nu_{i}(0)(1-(T/T_{\mathrm{N}})^{\alpha})^{\beta}$ (see
main text).
(b) Transverse relaxation rates $\lambda_{i}$ are seen to increase as
the magnetic transition is approached from below.
(c) The longitudinal relaxation rate $\Lambda$, which decreases as the
transition is approached from below.
\label{fitlt}}
\end{center}
\end{figure}

Across the measured temperature range, the two frequencies were found
to be 
in the proportions 
$\nu_{1}:\nu_{2} = 1:0.65$ while the relaxation rates 
$\lambda_{i}$ were found to be in the ratio $\lambda_{1}:\lambda_{2} =
1:0.75$. These quantities were fixed in these ratios during the
fitting procedure. The amplitudes and phases best describing the data
are given in \cite{fittable}. We note that the ratio of
longitudinal to transverse components is in excess of the
expected value of $A_{3}/(A_{1}+A_{2})=1/2$ (see below). Nonzero 
phases were also required to fit the data, as observed in
previous studies of Cu-based chain compounds of this sort \cite{tom}. 
The magnitudes of the frequencies were
fitted as a function of temperature with other parameters in 
Eq.(\ref{fitfunc}) fixed \cite{fittable}.
The resulting temperature evolution of the precession frequencies
is shown in Fig.\ref{fitlt}(a). 

From fits of $\nu_{i}$ to the form 
$\nu_{i}(T) =\nu_{i}(0)(1-T/T_{\mathrm{N}})^{\beta}$ close to
$T_{\mathrm{N}}$, 
we estimate $T_{\mathrm{N}}=107(1)$~mK, leading to a value of the
ratio $|k_{\mathrm{B}} T_{\mathrm{N}}/J|=0.0103(1)$. 
The parameter $\beta$ is highly sensitive to the chosen
value of $T_{\mathrm{N}}$ due to the difficulty of obtaining
data sufficiently close to the
transition. Using $T_{\mathrm{N}}=107(1)$~mK, we estimate 
$\beta = 0.18(5)$, which is consistent with
low dimensional (i.e.\ 2D Ising or 2D $XY$) behaviour.
Fits to 
$\nu_{i}(T) =\nu_{i}(0)(1-(T/T_{\mathrm{N}})^{\alpha})^{\beta}$
with $\alpha=3$ yield
$\nu_{1}(0)=1.922(4)$~MHz and $\nu_{2}(0)=1.257(3)$~MHz corresponding
to local magnetic fields at the two muon sites of $B_{1}=141.8(3)$~G
and $B_{2}=92.7(2)$~G. 

Further evidence for a magnetic transition is seen in the transverse 
relaxation rates $\lambda_{i}$, which increase as
the magnetic transition is approached from below (Fig.\ref{fitlt}(b)),
as would be expected
for critical slowing of fluctuations near a magnetic transition. 
The longitudinal relaxation rate 
$\Lambda$ decreases as the
transition is approached from below (Fig.\ref{fitlt}(c)), 
suggesting that this quantity is
dominated by the magnitude of the local magnetic field at the muon
site.

\begin{figure}
\begin{center}
\epsfig{file=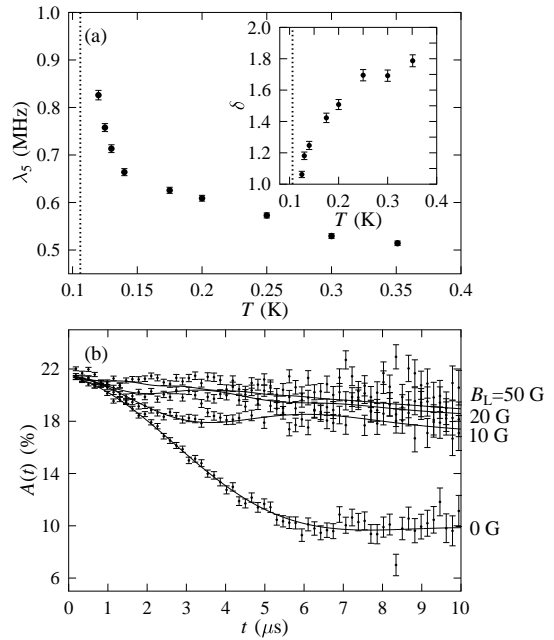,width=7.5cm}
\caption{
(a) Results of fitting data for $T>110$~mK to
Eq.(\ref{hightemp}). 
(Dashed lines show $T=T_{\mathrm{N}}$.)
{\it Inset:} The parameter $\delta$ changing from
$\delta=1$ (exponential relaxation) to $\delta=2$ (Gaussian relaxation)
as the dynamics move out of the muon time window with increasing
temperature. {\it Main panel:} The relaxation rate $\lambda_{5}$
increases
as the transition is approached from above.
(b) LF $\mu^{+}$SR spectra made at $T=2$~K in several applied
  magnetic
fields $B_{\mathrm{L}}$. Fits are shown to the LFKT function with $\Delta$ fixed at
$0.271$~MHz for all fields. \label{fitht}}
\end{center}
\end{figure}

At temperatures just above $T_{\mathrm{N}}$ the measured spectra are
best described by two components (Fig.\ref{data}(b)). 
One is a fast relaxing contribution,
with an amplitude $\sim 5$~\%,
likely to arise due to a fraction of the muons which initially
form radical states. This component will also exist in the
magnetically ordered regime, but is not resolvable from the 
longitudinal relaxation and partly explains why the value of $A_{3}$ 
is greater than that expected.
The second component observed in this regime is
an exponential functional form, characteristic of
fast dynamics in the local magnetic field at the muon sites \cite{hayano}. 
At higher temperatures still ($T > 400$~mK), this second component in the 
spectra appears Gaussian in form (Fig.\ref{data}(c)),
characteristic of 
the early time behaviour of a Kubo Toyabe  
function \cite{hayano}
due to the magnetic field distribution of a static, disordered system,
with slow dynamics preventing the characteristic recovery of the
polarization at late times. 
It is likely that at these elevated temperatures, 
the electronic moments are fluctuating too fast
compared to the muon time scale, and are therefore motionally narrowed
from the spectra. Instead the muon ensemble is depolarized by the
magnetic field distribution at the muon site due to the quasi-static nuclear
moments of the surrounding nuclei. This interpretation is supported by
the
observation that applied magnetic fields of magnitude $\geq 100$~G 
repolarize
the relaxation, as expected for a static distribution
(repolarization implies lim$_{t \rightarrow \infty} A(t) =
A(0)$).
The measured data for $T>T_{\mathrm{N}}$ were therefore fitted to the
function
\begin{equation}
A(t) = A_{4} \exp(-\lambda_{4} t) + A_{5} \exp(-\lambda_{5} t)^{\delta} +
A_{\mathrm{bg}}\label{hightemp}
\end{equation}
where the stretching parameter $\delta$ describes the passage from
exponential to Gaussian behaviour. The results of this analysis is
shown in Fig.\ref{fitht}(a). The fact that the relaxation rate $\lambda_{5}$
increases sharply as the proposed transition temperature is approached 
from above lends further weight to the case for the existence
of a magnetic transition. 

In order to further investigate the high temperature behaviour of the
material measurements were made in an applied longitudinal magnetic
field (LF) $B_{\mathrm{L}}$ using a $^{4}$He cryostat at the ISIS facility. 
The pulsed structure of the muon source at ISIS
has the advantage of allowing the observation of the muon polarization
at longer times than at a continuous source such as S$\mu$S. This
enables us to better characterise the data with the
Kubo Toyabe function generalised for applied LFs
\cite{hayano}.    
The loss of time resolution at a pulsed source causes the fast
relaxing
component described above to be missing from the signal entirely. 
The results of these measurements, 
made at $T=2$~K, are shown in Fig.\ref{fitht}(b).
The spectra are found to be approximately
temperature independent in the range $2 \leq T \leq 50$~K. The
application of a field repolarizes the slowly relaxing component of
asymmetry for $B <
100$~G  consistent with the interpretation that the relaxation at these
temperatures is due to the disordered, quasistatic nuclear magnetic
moments. 
The measured ZF and LF spectra for both materials at 2~K are well described
by the longitudinal field Kubo Toyabe (LFKT) function
describing the relaxation due to an ensemble of  disordered
magnetic moments, although the LFKT function must be multiplied by an
exponential function with constant relaxation
rate to account for the weak relaxation due to dynamic fluctuations \cite{francis}.
This treatment has only one free parameter
$\Delta=\gamma_{\mu}\sqrt {\langle B_{\mathrm{loc}}^{2} \rangle} $, where
$\langle B_{\mathrm{loc}}^{2} \rangle$ is the second moment of the
average local magnetic
field distribution at the muon sites in ZF. This analysis yields a value of
$\Delta=0.271$~MHz for CuPzN.

An efficacious
method for estimating the interchain coupling $J'$ in LCHAFM's
has recently been developed, based on a modified random phase
approximation, 
modelled with classical and quantum Monte Carlo simulations \cite{yasuda}.
This approach leads to an empirical formula for the interchain
coupling, given by
\begin{equation}
|J'|/k_{\mathrm{B}} = \frac{T_{\mathrm{N}}}
{4c 
\sqrt{ 
\ln( 
\frac{\lambda |J|}
{k_{\mathrm{B}}T_{\mathrm{N}}
})
+
\frac{1}{2} \ln \ln
( 
\frac{\lambda |J|}{k_{\mathrm{B}}T_{\mathrm{N}}
})}
},\label{jprime}
\end{equation}
with $\lambda = 2.6$ and $c=0.233$. Substituting our value of 
$T_{\mathrm{N}}=107(1)$~mK and using
$J/k_{\mathrm{B}}=-10.3(1)$~K, yields $|J'|/k_{\mathrm{B}} \simeq
0.046$~K and a ratio $|J'/J| = 4.4 \times 10^{-3}$. 
The parameters describing a number of copper-chain compounds are presented in
table \ref{chaintable}. 
Comparing values of the ratio $|J'/J|$ shows that CuPzN is 
approximately five times better isolated than copper bispyridine 
dichloride (Cu(py)$_{2}$Cl$_{2}$ or CPC) \cite{duffy,endoh}, 
while $|J'/J|$ for
CuPzN is around one quarter of that for the best isolated LCHAFM,
Sr$_{2}$CuO$_{3}$ (although this material has a QCP at a field
around 200 times larger than that for CuPzN) \cite{kojima,motoyama}.

\begin{table}
   \caption{Parameters for copper-chain
compounds Sr$_{2}$CuO$_{3}$ \cite{motoyama,kojima}, 
KCuF$_{3}$\cite{satija}, 
CPC(=Cu(py)$_{2}$Cl$_{2}$) \cite{duffy}
 and CuPzN (\cite{hammar} and this work)
with values of $|J'/J|$ estimated from Eq.(\ref{jprime}).\label{chaintable} }
     \begin{ruledtabular}
   \begin{tabular}{lllll} 
                 &  $|J|/k_{\mathrm{B}}$~(K)& $T_{\mathrm{N}}$~(K) & $|k_{\mathrm{B}}T_{\mathrm{N}}/J|$ & $|J'/J|$ \\\hline
Sr$_{2}$CuO$_{3}$ & 2200 & 5.4 &  $2.5 \times 10^{-3}$  & $9.3 \times 10^{-4}$\\
KCuF$_{3}$       &  406  &  39 & $ 9.6 \times 10^{-2}$ &  $5.2 \times 10^{-2}$\\
CPC              &   26  & 1.1 & $ 4.2 \times 10^{-2}$ & $2.1 \times 10^{-2}$ \\
CuPzN            &  10.3 & 0.107 & $1.0\times 10 ^{-2}$ & $4.4 \times 10^{-3}$\\
   \end{tabular}
     \end{ruledtabular}
\end{table}

The failure of specific heat studies \cite{mennenga} to detect
an anomaly
due to the onset of 3D LRO in CuPzN
down to 70~mK, may be attributed to the 
value of the ratio $|J'/J|$.
Investigations using a stochastic series quantum Monte Carlo
method \cite{sengupta} show that the height of the expected peak 
decreases with decreasing 
$|J'|$, becoming
nearly linear below $|J'/J|=0.2$. For smaller value of the 
ratio $|J'/J|$, the entropy change at the critical
temperature is so small that it makes an undetectable contribution to
the measured specific heat.

In conclusion we demonstrated the existence of long range
magnetic order in the LCHAFM material CuPzN with
$T_{\mathrm{N}}=107(1)$~mK. This leads to an estimate of the
interchain coupling of $|J'|/k_{\mathrm{B}}\simeq 0.046$~K. 
For many 1D systems, small values of $|J'/J|$ prevent 
the detection of 3D LRO with specific heat, and reduced ordered
moments make the use of other techniques problematical
(coupled chain mean-field theory \cite{schulz} then predicts an ordered moment size 
of
$\sim 0.05 \mu_{\mathrm{B}}$ for CuPzN).
Thus we expect that muons may provide a particularly valuable
tool for future magnetic measurements of the key parameters
of highly anisotropic low-dimensional systems. 

We thank W. Hayes for useful discussion. 
Part of this work was carried out at the Swiss Muon Source, 
Paul Scherrer Institute, Villigen, Switzerland and at the ISIS
facility, Rutherford Appelton Laboratory, UK.
This work is supported by the EPSRC, UK. 
T.L acknowledges support from the European Commission under the 6th 
Framework Programme through the Key Action: Strengthening the European 
Research Area, Research Infrastructures. Contract no: RII3-CT-2003-505925


\begin{thebibliography}{xx}
\bibitem{muller}G. M\"{u}ller {\it et al.}, Phys. Rev. B, {\bf 24}, 1429
  (1981). 
\bibitem{karbach}
M. Karbach {\it et al.} Phys. Rev. B, {\bf 55}, 12510 (1997).
\bibitem{sachdev} S. Sachdev, {\it Quantum Phase Transitions}
(Cambridge University Press, Cambridge, England, 2000).
\bibitem{bogoliubov}
N.M. Bogoliubov, {\it et al.}, Nucl. Phys. B{\bf 275}, 687 (1986)
\bibitem{fledderjohann}
A. Fledderjohann {\it et al.}, Phys. Rev. B {\bf 54} 7168 (1996).
\bibitem{hammar} 
P.R. Hammar {\it et al.}, Phys. Rev. B, {\bf 59} 1008 (1999).
\bibitem{villain}
J. Villain and J.M. Loveluck, J. Phys (France) Lett. {\bf 38},
L77 (1977).
\bibitem{stone}
M.B. Stone {\it et al.}, 
Phys. Rev. Lett., {\bf 91}, 37205 (2003).  
\bibitem{lefmann}
K. Lefmann and C. Rischel, Phys. Rev. B {\bf 54}, 6340 (1996);
\bibitem{karbach2}
M. Karbach and G. M\"{u}ller, Phys. Rev. B {\bf 62}, 14871 (2000); 
\bibitem{karbach3}
M. Karbach {\it et al.}, Phys. Rev. B {\bf 66}, 054405 (2002).
\bibitem{mennenga} G. Menenga {\it et al.}
J. Magn. Magn. Mater., {\bf 44}, 89 (1984).
\bibitem{steve}
S.J. Blundell, Contemp. Phys. {\bf 40}, 175 (1999).
\bibitem{tom}
T. Lancaster {\it et al.}
 J. Phys. Condens. Matter {\bf 16}, S4563 (2004).
\bibitem{fittable}
The fixed fitting parameters for Eq.(\ref{fitfunc}) best describing
the data measured below $T=110$~mK were found to be
$A_{1}=1.56$~\%, $\phi_{1} = -37.8^{\circ}$,
$A_{2}=0.63$~\%, $\phi_{2} = -73.5^{\circ}$,
and $A_{3}=10.42$~\%. 
\bibitem{hayano}
R.S. Hayano {\it et al.}  Phys. Rev. B {\bf 20}, 850 (1979).
\bibitem{francis} The relaxation function multiplied by
the LFKT function may be used to extract information on
spin-wave dynamics in these materials; F.L. Pratt {\it et al.}, in preparation.
\bibitem{yasuda}
C. Yasuda {\it et al.}, Phys. Rev. Lett., {\bf 94}, 217201 (2005).
\bibitem{duffy}
W. Duffy Jr., Phys. Rev. B, {\bf 9}, 2220 (1974).
\bibitem{endoh}
Y. Endoh {\it et al.}, Phys. Rev. Lett. {\bf 32}, 170 (1974).
\bibitem{kojima} 
K.M. Kojima {\it et al.}, Phys. Rev. Lett., {\bf 78},
1787 (1997).
\bibitem{motoyama}
N. Motoyama {\it et al.}, Phys. Rev. Lett., {\bf 76}, 3212 (1996).
\bibitem{satija}
 S.K. Satija {\it et al.} Phys. Rev.B {\bf 21},  2001 (1980).
\bibitem{schulz}
H.J. Schulz,  Phys. Rev. Lett. {\bf 77}, 2790 (1996).
\bibitem{sengupta}
P. Sengupta {\it et al.} Phys. Rev. B, {\bf 68}, 094423 (2003).
\end{thebibliography}
\end{document}